\documentclass[prb,twocolumn,showpacs,floatfix,amsmath,amssymb,superscriptaddress]{revtex4-2}
\usepackage{amsfonts}
\usepackage{stmaryrd}
\usepackage{bbm}
\usepackage{mathrsfs}
\usepackage{tipa}
\usepackage{amssymb}
\usepackage{txfonts}
\usepackage{graphicx}
\usepackage{dcolumn}
\usepackage{epstopdf}
\usepackage[colorlinks,linkcolor=blue,urlcolor=black,citecolor=blue]{hyperref}
\usepackage{multirow}
\usepackage{subfigure}
\usepackage{url}
\usepackage{upgreek}
\usepackage[utf8]{inputenc}
\usepackage[english]{babel}

\begin{document}

\newcommand*{\cm}{cm$^{-1}$\,}
\newcommand*{\Tc}{T$_c$\,}

\title{Ultrafast dynamics of optically excited charge carriers in the room-temperature antiferromagnetic semiconductor $\alpha $-MnTe}

\author{Changqing Zhu}
\author{Patrick Pilch}
\author{Anneke Reinold}
\author{Dennis Kudlacik}
\affiliation{Department of Physics, TU Dortmund University, 44227 Dortmund, Germany}
\author{Gunther~Springholz}
\author{Alberta Bonanni}
\affiliation{Institute of Semiconductor and Solid State Physics, Johannes Kepler University Linz, 4040 Linz, Austria}
\author{Marc Assmann}
\author{Mirko Cinchetti}
\author{Zhe Wang}
\affiliation{Department of Physics, TU Dortmund University, 44227 Dortmund, Germany}

\date{\today}

\begin{abstract}
We report on time-resolved optical and terahertz ultrafast  spectroscopy of charge-carrier dynamics in the room-temperature antiferromagnetic semiconductor $\alpha $-MnTe.
By optically pumping the system with 1.55 eV photons at room temperature, we excite charge carriers in the conduction band through the indirect band gap and investigate dynamical response of the nonequilibrium states using optical as well as terahertz transmission probes.
Three relaxation processes are revealed by their characteristic relaxation times of the order of 1, 10, and 100~ps, whose exact values are function of the pump fluence. For high pump fluences nonlinear dependence on the pump fluence is observed both in the optical and terahertz probes.
\end{abstract}

\maketitle

\section{Introduction}
Semiconductors with long-range antiferromagnetic order and zero net magnetic moment are promising versatile candidates for applications in spintronics \cite{jungwirth2016antiferromagnetic,baltz2018antiferromagnetic}. Especially for opto-spintronic applications the potential of antiferromagnetic semiconductors has not been fully investigated so far \cite{nvemec2018antiferromagnetic}. 
Hexagonal $\alpha$-MnTe is a representative antiferromagnetic semiconductor which can host a long-range antiferromagnetic order at room temperature because of its relatively high Néel temperature $T_N=307$~K \cite{Kelley1939}. A wide range of interesting magnetic properties and related functionalities have been reported recently in this material (see e.g. \cite{kriegner2016multiple,He2018,Zheng19,
bossini2020exchange,bossini2021femtosecond,Wang2021,
PhysRevLett.130.036702,moseley2022giant,Yao2022}).

The hexagonal crystalline structure of $\alpha$-MnTe with a space group of $P6_3/mmc$ is constituted by alternating hexagonal layers of Mn and Te atoms \cite{EFREMDSA2005267,moseley2022giant}.
Below $T_N$ the Mn spins within the hexagonal layers are aligned ferromagnetically, while  along the c axis these layers are coupled antiferromagnetically  \cite{Kunitomi64}.
The magnetic phase transition at $T_N$ manifests not only as a peak in specific heat \cite{Kelley1939}, but also as a pronounced change of electric resistivity and thermopower \cite{Zheng19}. Below $T_N$ anisotropic magnetoresistance and anomalous Hall effect have been reported \cite{kriegner2016multiple,PhysRevLett.130.036702}, which also indicate coupling between magnetic and charge degrees of freedom. Moreover, coupling between spin and lattice was recently evidenced by optical spectroscopic studies \citep{bossini2020exchange,bossini2021femtosecond}.
By optically pumping the system with a 2.4~eV laser pulse the magnetic response was probed by measuring time-resolved polarization rotation of another laser pulse of 1.72~eV \cite{bossini2021femtosecond}. While the change of polarization can be associated with the magneto-optic Faraday effect, an additional oscillatory behavior at a frequency of 5.3 THz was found from the probe beam \cite{bossini2021femtosecond}, which corresponds to an optical phonon mode in $\alpha $-MnTe \cite{onari1974temperature}.

The experimentally determined indirect band gap in $\alpha $-MnTe is about 1.27--1.46~eV depending on temperature \cite{Ferrer00,kriegner2016multiple,bossini2020exchange}. This value is consistent with ab initio band structure calculations \cite{Wei87,PhysRevLett.130.036702}.
In the fabrication of heterostructures, $\alpha $-MnTe has been used as exchange bias for topological insulators and metals \cite{He2018,Wang2021}, and was also found to affect the properties of superconductors \cite{Yao2022}.
The magnetic, electric, and thermal properties of $\alpha $-MnTe are sensitive to chemical doping \cite{moseley2022giant} and external conditions \cite{Przezdziecka05}. For example, by 5\% Li doping the spin order is tuned from in-plane to c-axis orientation   \cite{moseley2022giant}. At the same time, thermopower and electric resistivity are reduced by a factor of more than 50\% and 70\%, respectively \cite{Zheng19}. Comparing with single crystals, $\alpha $-MnTe thin films grown on a substrate can exhibit a lower Néel temperature and a larger band gap \cite{Przezdziecka05}.  

In this work we investigate the ultrafast dynamics of optically excited charge carriers above the band gap in $\alpha $-MnTe at room temperature using both terahertz and optical probe beams. Relaxation processes are revealed by their characteristic relaxation time scales. While a linear dependence of the optical and terahertz response on pump fluence is found in the low-fluence limit, we observe a clear nonlinear dependence at high fluences. Our studies provide a systematic characterization of the ultrafast charge-carrier dynamics in $\alpha $-MnTe.

\section{Experimental details}

\begin{figure}[b]
\centering 
\includegraphics[width=\linewidth]{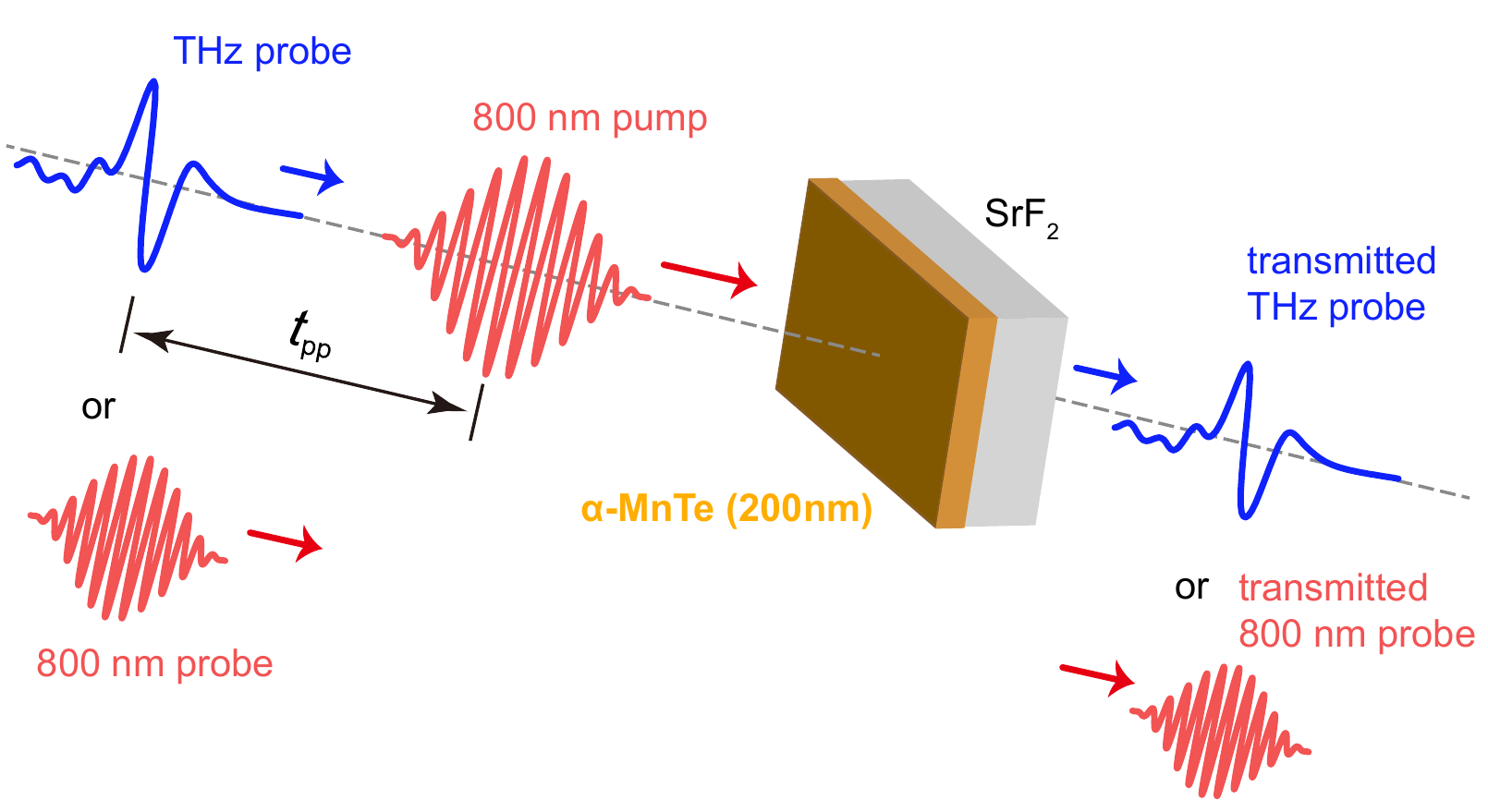}
\caption{Schematic of 800 nm-pump 800 nm-probe or THz probe spectroscopy. The pump and probe pulses are synchronized with a variable delay time $t_\text{pp}$. The measured sample is an $\alpha$-MnTe thin film grown on a SrF$_2$ substrate.}
\label{Fig.1} 
\end{figure}

In our time-resolved pump-probe spectroscopic experiment, 800 nm laser pulses (corresponding to 1.55 eV in photon energy) from a Ti:Sapphire laser system (1~kHz, 100~fs) are used to excite the $\alpha $-MnTe sample (see Fig.~\ref{Fig.1} for an illustration). To probe the response of the optically excited states, we synchronize terahertz or 800 nm pulses with the pump pulses.
Transmitted terahertz field or 800 nm intensity is recorded as a function of the tunable time delay $t_\text{pp}$ between the pump and probe pulses. 
The terahertz probe pulses are generated using a LiNbO$_{3}$ crystal \cite{Hebling02} and gate-detected by electro-optic sampling with a 0.5 mm thick $<$110$>$-cut ZnTe crystal \cite{Planken01}.
The transmission of the 800 nm probe beam is measured by using a balanced photodetector. 
For the measurements an $\alpha $-MnTe thin-film sample with a thickness of 200 nm is epitaxially grown on a (111) oriented SrF$_{2}$ substrate. All measurements are performed at room temperature.

\begin{figure}[t]
\centering
\includegraphics[width=\linewidth]{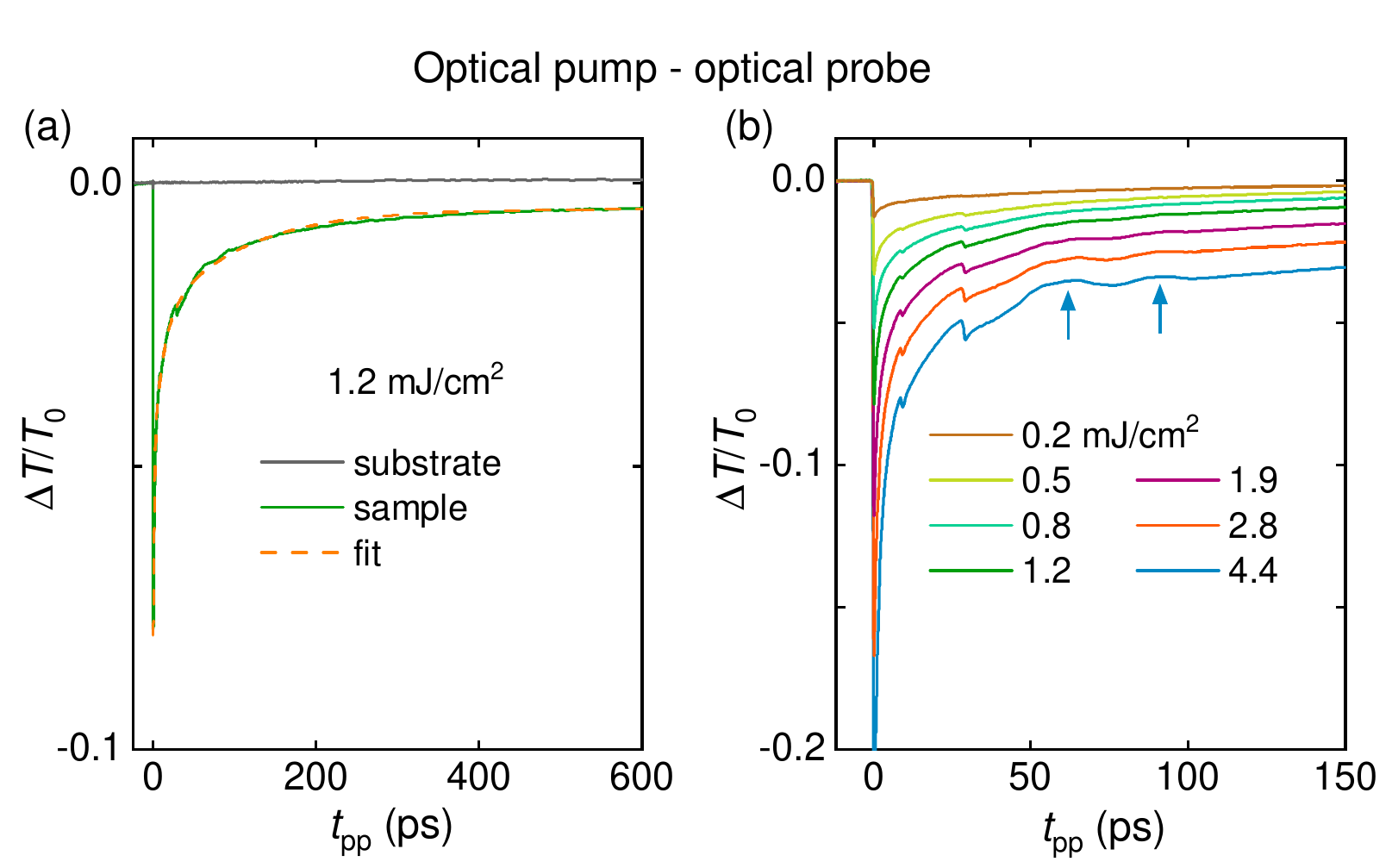}
\caption{(a) Transient optical transmission $ \triangle T/T_{0} $ through a bare SrF$_{2}$ substrate and a $\alpha $-MnTe film on substrate as a function of delay time $t_{\text{pp}} $ between the optical-pump and optical-probe pulses. $T _\text{0}$ is the static optical transmission without photoexcitation. 
The dashed line represents a fit of three exponential decay functions, whose fitting parameters are presented in Fig.~\ref{Fig.6}.
(b) $\triangle T/T_{0}$ of the sample for various pump fluences. The two arrows for 4.4~mJ/cm$^2$ indicate an oscillatory behavior, which is observed at the highest fluences, corresponding to a frequency of 35~GHz.}
\label{Fig.2} 
\end{figure}

\section{Experimental results}

\subsection{Optical-pump optical-probe measurements}
In order to study the relaxation dynamics of the excited charge carriers in $\alpha$-MnTe, we first use 800~nm laser pulses to probe the material after an excitation by 800~nm pulses.
Fig.~\ref{Fig.2}(a) displays the pump induced relative changes of transmission $\Delta T/T_0$ for a pump fluence of 1.2~mJ/cm$^{2}$ through $\alpha$-MnTe on a SrF$_{2}$ substrate, and through a bare substrate for reference.
The bare substrate does not exhibit a detectable pump-induced response, therefore the pump-probe response obtained in our experiment is primarily from the $\alpha $-MnTe thin film.
Due to the optical pump, the transmission at 800~nm drops by about 10\% at zero pump-probe delay $t_\text{pp}=0$~ps, which is followed by a continuous increase of the transmission towards the value of the equilibrium state.

The results for various pump fluences are presented in Fig.~\ref{Fig.2}(b). With increasing pump fluence the pump induced changes are enhanced and the relaxation of the system becomes slower. For higher fluences than 1.9~mJ/cm$^{2}$, one can clearly see oscillation-like behavior, as marked by the two arrows, which damps out after 100~ps.
For $t_\text{pp}<50$~ps, the two kinks around 10 ps and 30 ps are caused by a back-reflection of the laser pulse in the sample and by a parasitic post-pulse from the laser system, respectively. Both of them blur the observation of additional oscillation maxima at smaller $t_\text{pp}$. Since the time difference between the two observed maxima is about 28.3~ps, corresponding to a rather low frequency of 35~GHz, the oscillation-like behavior is likely due to acoustic phonons excited by the impact of the high laser fluence.

\subsection{Optical-pump terahertz-probe measurements}

\begin{figure}[t]
\centering
\includegraphics[width=\linewidth]{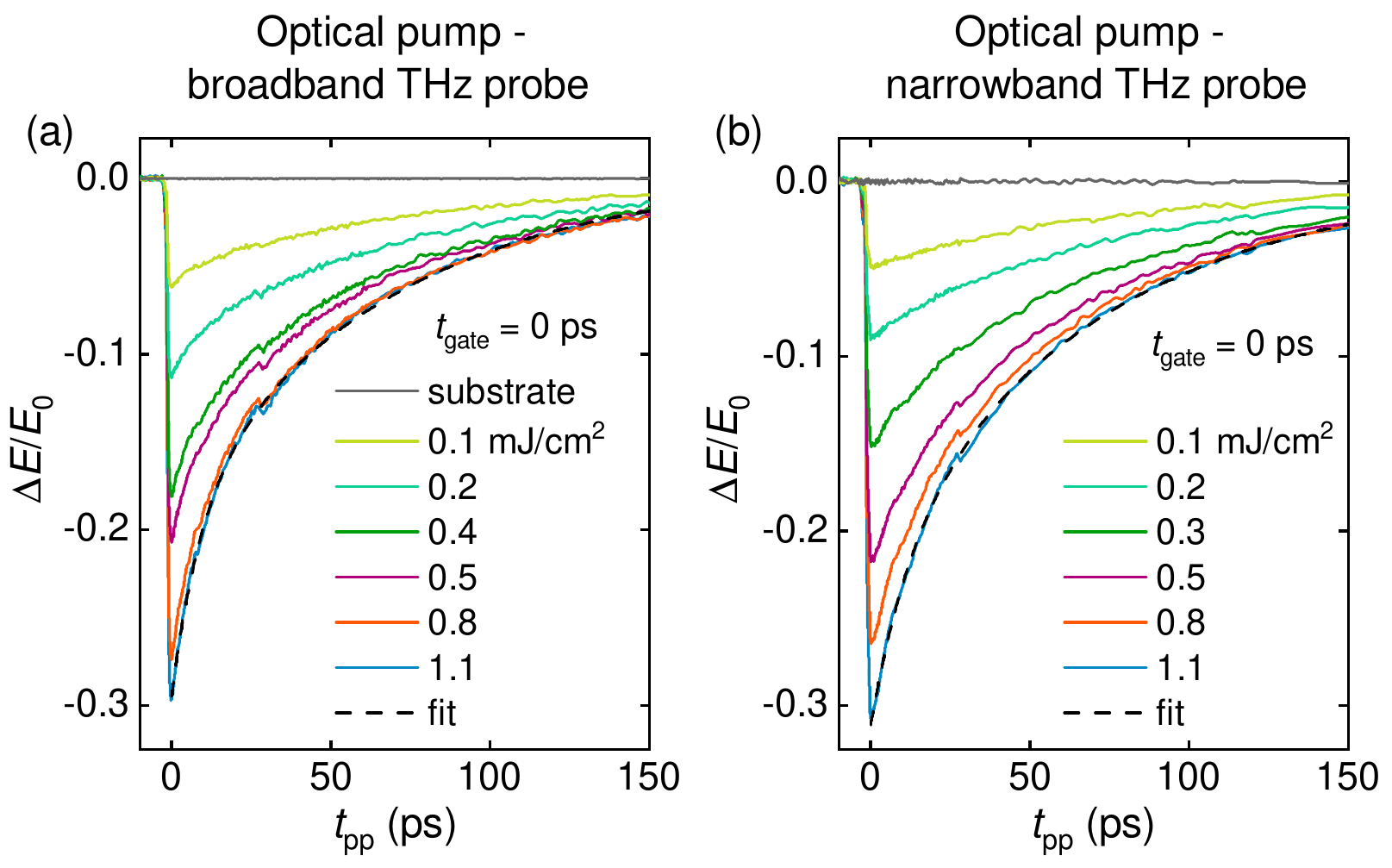}
\caption{(a) Broadband and (b) narrowband THz probe transmission of the optically excited states for various pump fluences. The pump induced changes of the transmitted THz peak electric field $\Delta E/E_0$ are plotted versus the pump-probe delay $t_\text{pp}$.
$E_{0}$ is the transmitted peak electric field without pump.  
The dashed lines represent fits of double exponential functions (see Fig.~\ref{Fig.7} for the fitting parameters).  
The peak fields of broad- and narrowband probes are 140 and 130~kV/cm, respectively.
 }
\label{Fig.3}
\end{figure}

Since the energy of terahertz photons is rather low (1~THz $\sim$ 4~meV) in comparison with 800~nm (1.55~eV) photons, the terahertz photons cannot induce interband transitions but are more sensitive to intraband dynamics \cite{Reinhoffer20,Kovalev20,Kovalev21,Germanskiy22,Reinhoffer22}. Thus, by using terahertz probe beams we can provide complementary information on the ultrafast charge carrier dynamics in $\alpha$-MnTe.
To comprehensively characterize the nonequilibrium dynamical response, we use both single-cycle and multi-cycle THz pulses, which are broad- and narrowband in frequency domain, respectively, to probe the optically excited states.

\begin{figure}[t]
\centering
\includegraphics[width=\linewidth]{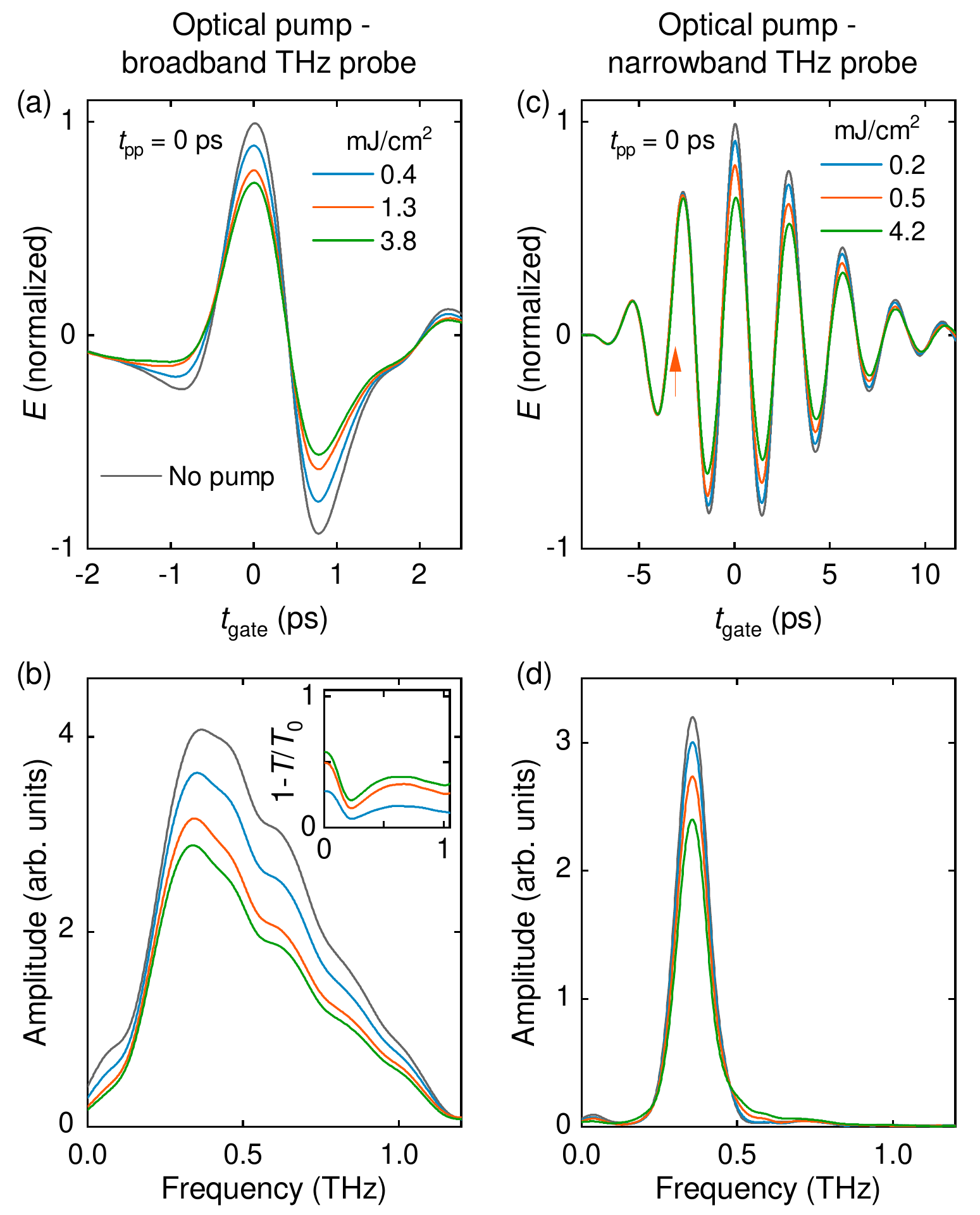}
\caption{Waveforms of transmitted THz electric fields measured at $t_\text{pp}=0$~ps in time domain for (a) the broad- and (c) narrowband THz probe pulses at various pump fluences. The corresponding spectra in frequency domain for (b) the broad- and (d) narrowband THz probe pulses. 
The inset in (b) presents the pump-induced transmission change ($1- T/T_{0} $) for various pump fluences, where $T_{0}$ denotes the transmission without pump.
The arrow in (c) indicates the onset of pump-induced transmission change [cf. Fig.~\ref{Fig.3}(b)].   }
\label{Fig.4}
\end{figure}

Figure~\ref{Fig.3}(a) and \ref{Fig.3}(b) display the optically induced changes $\Delta E/E_0$ of the transmitted terahertz electric field for single-cycle and multi-cycle THz pulses, respectively, at a gate time delay $t_\text{gate}=0$~ps, which means the gate laser pulse overlap with the peak-field position of the THz pulses in time.
In the frequency domain, the spectrum of the broadband probe beam extends from about 0.1 to 1 THz [see Fig.~\ref{Fig.4}(b)], while the narrowband spectrum is centered around 0.35~THz as a corresponding bandpass filter with 20\% FWHM is used [see Fig.~\ref{Fig.4}(d)]. 
The pump induced changes are qualitatively similar for the broad- and narrowband probes. The changes are negative, corresponding to a reduced transmission of the THz pulses due to the photoexcitation. For a pump fluence around 1~mJ/cm$^{2}$, the maximum change $\Delta E/E_0$ of the probe pulses can be as high as 30\%.

\begin{figure}[t]
\centering
\includegraphics[width=\linewidth]{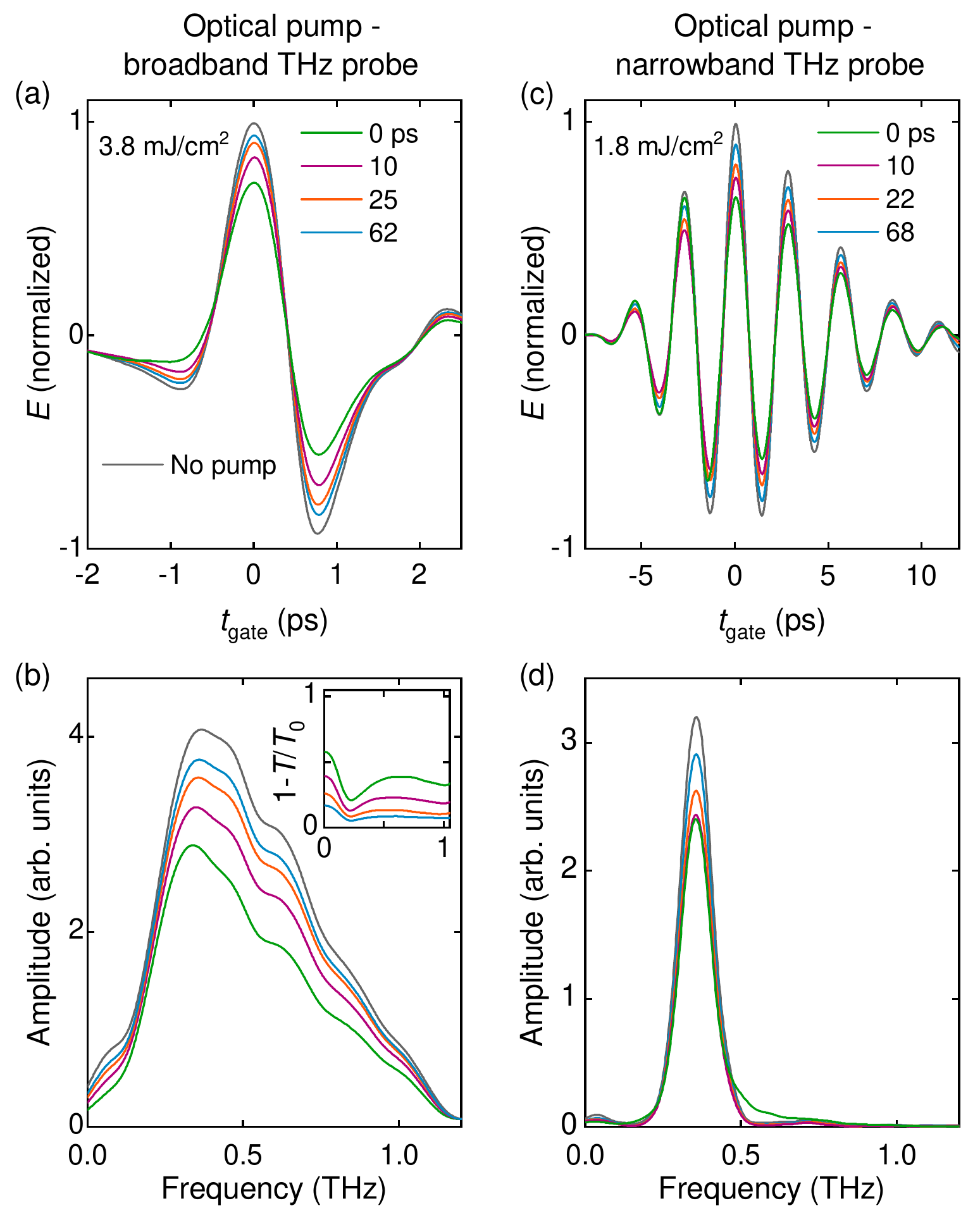}
\caption{Waveform of transmitted THz electric fields measured at different pump-probe delays $t_\text{pp}$'s in time domain for (a) the broad- and (c) narrowband THz probe pulses with pump fluences of 3.8 and 1.8 mJ/cm$^{2}$, respectively. (b) and (d) present the corresponding spectra in frequency domain.  
The inset in (b) presents the pump-induced transmission change ($1- T/T_{0} $) for various pump-probe delays, where $T_{0}$ denotes the transmission without pump.}
\label{Fig.5}
\end{figure}

\begin{figure*}[t]
\centering
\includegraphics[width=17cm]{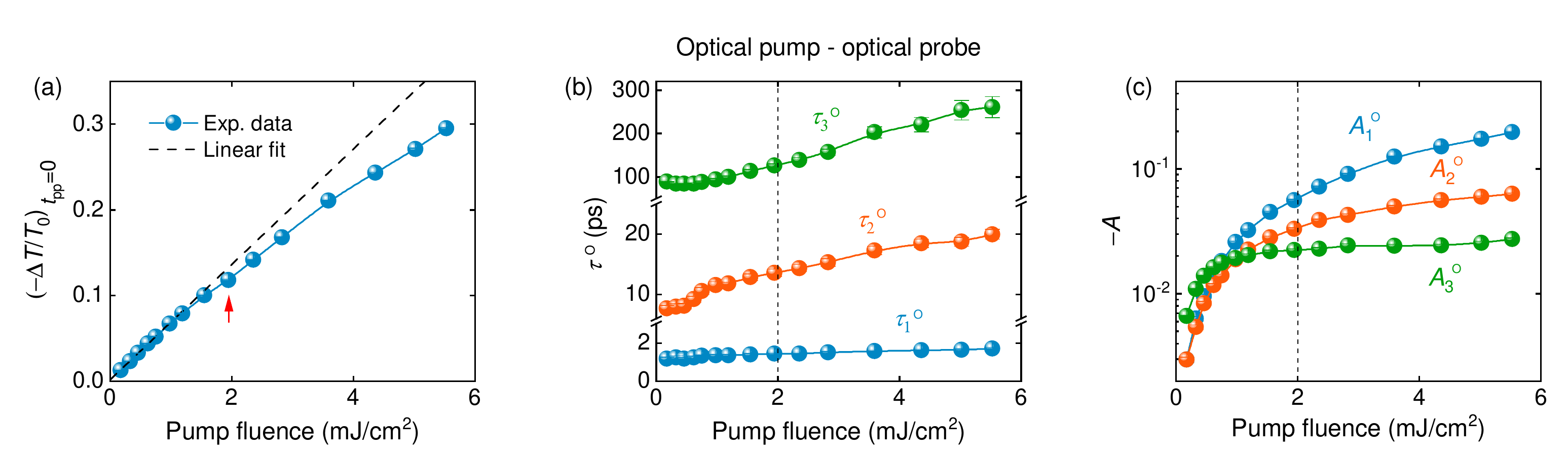}
\caption{(a) Pump fluence dependence of transient optical transmission $ (-\triangle T/T_{0})_{t_{\text{pp}} = 0} $. The dashed line is a linear fit to the lowest fluences. The arrow indicates the onset of a deviation from the linear dependence. (b) The obtained relaxation times $\tau^O_{1}$, $\tau^O_{2}$, $\tau^O_{3}$ and (c) the corresponding amplitudes $A^O_{1}$, $A^O_{2}$, $A^O_{3}$  for fitting the pump-probe signal with a sum of three exponential decay functions (see Fig.~\ref{Fig.2}). }  
\label{Fig.6}
\end{figure*}

The pump-induced changes of the terahertz response can be further characterized by measuring of the transmitted terahertz waveforms in the time domain, which are shown in Fig.~\ref{Fig.4}(a) and \ref{Fig.4}(c) for selected pump fluences and $t_\text{pp}=0$~ps. The pump-probe time delay of $t_\text{pp}=0$~ps is chosen corresponding to the maximum change induced by the pump pulse (see Fig.~\ref{Fig.3}).
Fourier transformation of the time-domain traces provides the spectra in frequency domain,
which are presented in Fig.~\ref{Fig.4}(b) and \ref{Fig.4}(d), respectively, for the broad- and narrowband probes.
After the optical excitation the system becomes more absorptive not only for the narrowband probe but in the whole available terahertz frequency range.
The optically induced absorption is already very evident for a relatively small pump fluence, e.g. 0.2 mJ/cm$^{2}$ in Fig.~\ref{Fig.4}(d).
With increasing fluence the changes are however not linear, a quantitative discussion of which will be provided below (see Fig.~\ref{Fig.7}).  {In the inset of Fig. \ref{Fig.4}(b), we present the pump-induced transmission changes $(1-T/T_0)$ in frequency domain for various pump fluences, where $T_0$ denotes the transmission without pump.
The increase towards the low-frequency limit is expected as a Drude-type metallic response of optically excited charge carriers. Such response is stronger for higher pump fluence, in line with the enhanced population of photoinduced charge carriers. We shall note that for a far-from-equilibrium system, the Drude model may not necessarily provide a valid description of a metallic response.

Furthermore, to probe the relaxation dynamics we measure the changes of the terahertz waveforms at different pump-probe delays after $t_\text{pp}=0$~ps, which are summarized in Fig.~\ref{Fig.5}.
For a fixed pump fluence of e.g. 3.8~mJ/cm$^{2}$ [Fig.~\ref{Fig.5}(a)(b)], the pump-induced changes reduce monotonically with increasing pump-probe delay $t_\text{pp}$ for the whole THz waveform in time domain and also for the whole spectral range in frequency domain. A similar evolution of the transmitted terahertz waveform with pump-probe delay is observed for the narrowband probe [Fig.~\ref{Fig.5}(c)(d)]. As reflected by time-dependence evolution of $(1-T/T_0)$ in the inset of Fig.~\ref{Fig.5}(b) the Drude-type metallic response decreases with increasing pump-probe delay. By comparing Fig.~\ref{Fig.5} and Fig.~\ref{Fig.4} phenomenologically, one can see that the effects of reducing pump fluence are similar to that of increasing pump-probe delay. This qualitatively reflects the fact that both variations lead to a decrease of the optically excited charge carriers in the nonequilbrium states. However, such qualitative observation is valid essentially only in the linear regime. As we will discuss below, a nonlinear dependence regime can be found by a quantitative analysis of the experimental data.

\section{Discussion}

For the bare SrF$_{2}$-substrate under the 800 nm pump, both the 800 nm and the THz responses are essentially zero (see Fig.~\ref{Fig.2} and Fig.~\ref{Fig.3}). These are consistent with the fact that the pump photon energy of 1.55~eV is much smaller than the band gap of about 9--10 eV in SrF$_{2}$ \cite{Myasnikova20}. The observed pump-probe signals reflect the dynamical response of the optically excited charge carriers in the $\alpha $-MnTe thin films.

To quantify the 800 nm response of the nonequilibrium states, we plot the optically induced maximum change of the 800 nm transmission versus pump fluence in Fig.~\ref{Fig.6}(a). For low pump fluences we can fit the fluence dependence by a linear function, as shown by the dashed line in Fig.~\ref{Fig.6}(a)]. However, above $ \sim $ 2~mJ/cm$^{2}$, it clearly shows that the experimental data deviate evidently from the linear dependence.

To provide a further quantitative analysis of the experimental data, we simulate the relaxation behavior of the optically induced changes after $t_\text{pp}=0$ ps by a sum of the exponential functions  $A_i\exp(-t/\tau_i)$,
where $\tau_i$ represents the characteristic relaxation time of different relaxation processes and $A_i$ is the corresponding amplitude. For the 800~nm (Fig.~\ref{Fig.2}) and THz probes (Fig.~\ref{Fig.3}), $A_i$ denotes 800~nm transmission and transmitted THz electric field, respectively.
We find that the relaxation behavior probed by 800 nm transmission and THz transmission should be simulated by three and two relaxation functions, respectively.

A representative fit by three exponential functions to the 800 nm probe data is shown in Fig.~\ref{Fig.2}(a). The obtained three relaxation times $\tau^O_1$, $\tau^O_2$, and $\tau^O_3$ are presented in Fig.~\ref{Fig.6}(b) as a function of pump fluences.
The three relaxation processes are characterized by very different relaxation times, which are of the order of $\tau^O_1=1$~ps, $\tau^O_2=10$~ps, and $\tau^O_3=100$~ps, respectively, at the lowest fluences. With increasing fluence, they increase continuously to about 2, 20, and 300~ps, respectively, at about 6~mJ/cm$^{2}$.
The corresponding amplitudes $A^O_{1}$, $A^O_{2}$ and $A^O_{3}$ obtained from the fits are presented in Fig.~\ref{Fig.6}(c). It evidently shows that the faster decaying process contributes a more dominant amplitude change.
We should note that at the highest fluences due to the oscillation behavior [see Fig.~\ref{Fig.2}(b)] the fit is rather a parameterization of the data. Although the obtained values of the relaxation times have larger uncertainty, they are still different by orders of magnitude, which clearly distinguish the three relaxation processes.
The obtained relaxation times $\tau^O_1$, $\tau^O_2$, and $\tau^O_3$ are characteristic for processes involving electron-phonon scattering, phonon-phonon scattering, and heat diffusion, respectively. 
 While quantitatively these values are consistent with those reported from 2.4~eV-pump 1.7~eV-probe measurements~\citep{bossini2021femtosecond}, we note that the THz pulses probe intraband processes whereas the 1.7~eV probe involved rather interband transitions.

\begin{figure}[t]
\centering
\includegraphics[width=\linewidth]{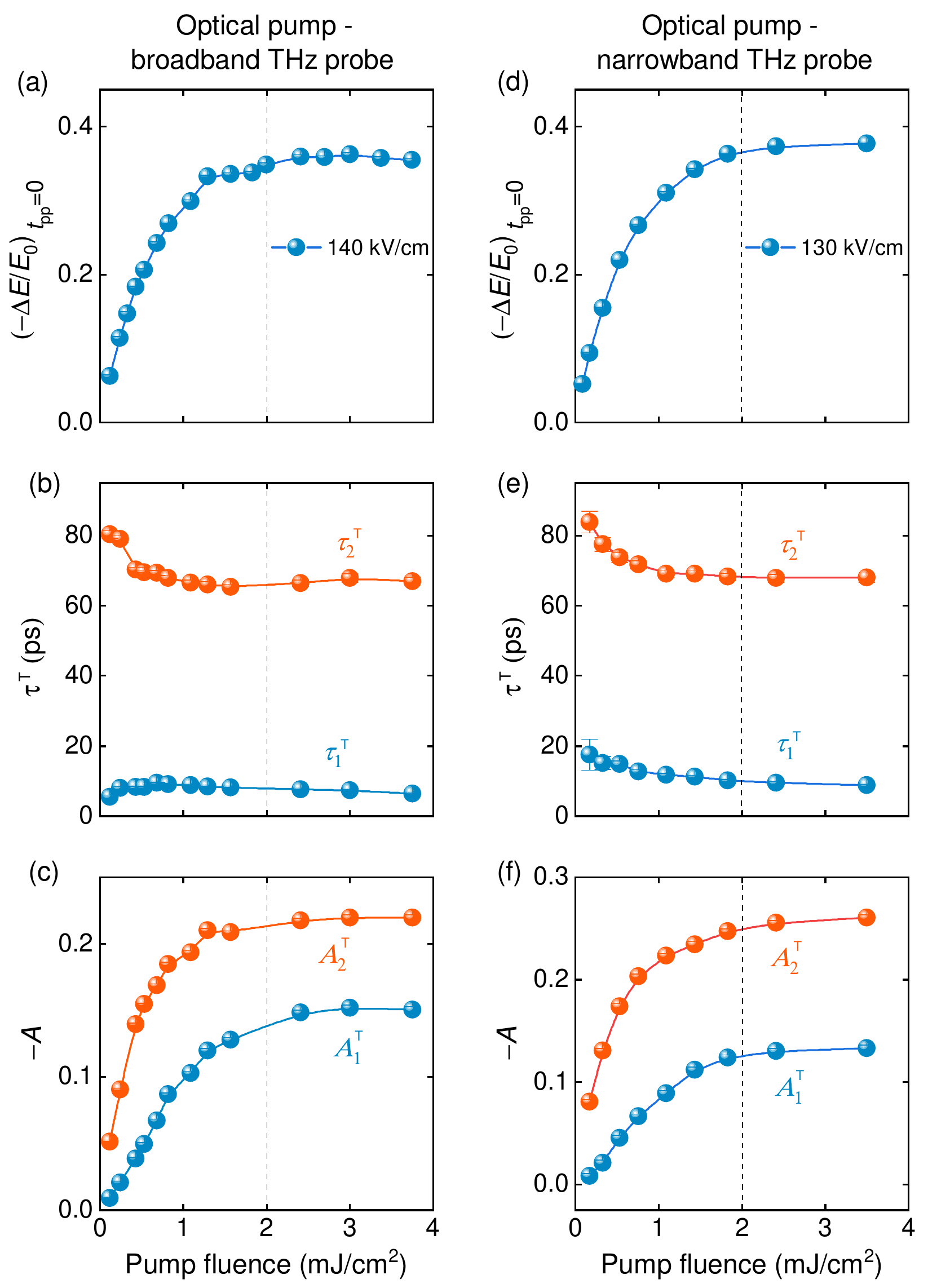}
\caption{Optical pump induced changes of the transmitted THz electric field $ (-\triangle E/E_{0})_{t_{\text{pp}}=0}$ as a function of pump fluence for (a) the broadband and (b) the narrowband THz probes. The peak electric fields of the THz probe pulses are 140 and 130~kV/cm, respectively.  (c)(d) The obtained relaxation times $\tau^T_{1}$, $\tau^T_{2}$ and (e)(f) amplitudes $A^T_{1}$, $A^T_{2}$ by fitting the pump-probe signals with two exponential functions (see Fig.~\ref{Fig.3}).}
\label{Fig.7}
\end{figure}

In contrast to the optical probe, we find that our terahertz probe is sensitive to only the two slower relaxation processes. As shown in Fig.~\ref{Fig.3}, the relaxation behaviors can be fitted by a sum of two exponential relaxation functions, whose characteristic relaxation times are summarized in Fig.~\ref{Fig.7}(b) and \ref{Fig.7}(e) for the broad- and narrowband THz probes, respectively, with the corresponding amplitudes presented in Fig.~\ref{Fig.7}(c) and \ref{Fig.7}(f).  
For the two different probes, the obtained values of the relaxation times are similar. At the lowest fluence $\tau^T_2$ is about 80~ps which decreases slightly with increasing fluence and levels off above 2~mJ/cm$^{2}$.
The values of $\tau^T_1$ are about $\sim 10$~ps at high fluences, whereas at the lowest fluence they are slightly different for the broad- and the narrowband probe beams where the fit uncertainty is also larger due to the smaller signals.
 
The level-off of $\tau^T_1$ and $\tau^T_2$ occurs not only concomitantly with the appearance of the nonlinear dependence observed in the 800 nm probe [see Fig.~\ref{Fig.6}(a)], but is also accompanied by a saturation-like feature in the terahertz response. As shown in Fig.~\ref{Fig.7}(a) and ~\ref{Fig.7}(d), the pump induced maximum change of the transmitted electric field exhibits a saturation above the pump fluence of 2~mJ/cm$^{2}$, while below this fluence the dependence increases monotonically.

For $\alpha$-MnTe the pump photon energy of 1.55~eV is just above the indirect band gap of 1.46~eV. The observed reduction of 800 nm and THz transmission is clear evidence of the optically excited charge carriers. Phenomenologically, the dependences of the THz probe on the pump fluence (Fig.~\ref{Fig.4}) and on the pump-probe delay (Fig.~\ref{Fig.5}) look very similar.
At lower pump fluences the THz transmission is relatively higher (Fig.~\ref{Fig.4}), because fewer charge carriers are excited. Also for longer pump-probe delays the THz transmission is relatively higher (Fig.~\ref{Fig.5}), because an increased portion of the excited charge carriers have relaxed already. These qualitative dependencies hold for both the broad- and narrowband probes, at least for the available THz spectral range. 

One possible reason for the observed nonlinear behavior in the 800 nm probe and in the terahertz probe could be the Pauli repulsion of the excited electrons. This effect may become evident only for sufficiently high density of optically excited charge carriers at high pump fluences.
At the same time, for this indirect bandgap semiconductor, phonons should be involved to assist the interband excitations, therefore the available density of states for the involved phonons could be another limiting factor, which might lead to the observed nonlinear dependence at high fluences. 

Since our experiment was carried out at a room temperature of 293~K, not far below $T_N=307$~K of a bulk $\alpha$-MnTe crystal \cite{Kelley1939}, the magnetic moments remain largely disordered. For an $\alpha$-MnTe thin film previous studies showed that anisotropic magnetoresistance disappears already at 285~K \cite{kriegner2016multiple}, suggesting that $T_N$ in a thin film may be even lower than 285~K. Therefore, in the nonequilibrium dynamics we probed the majority of the magnetic degrees of freedom is essentially not involved. This means that to investigate magnetic dynamics, it is necessary to perform the measurements at much lower temperatures where the sublattice magnetization is strongly enhanced (see e.g. \cite{Przezdziecka05}).

In contrast to earlier time-resolved spectroscopic studies using a 2.4~eV pump pulse \cite{bossini2021femtosecond} which excited an optical phonon mode of about 5~THz ($\sim 20$~meV) \cite{onari1974temperature}, we resolve rather an acoustic phonon-like feature at 35~GHz. The difference may result from the fact that the energy of the 2.4~eV photons was sufficiently high to allow various possible interband excitation mechanisms, whereas the photon energy of 1.55~eV is only slightly larger than the indirect band gap, which cannot trigger the dynamical processes involving the 5~THz optical phonons.

To conclude, we have studied the ultrafast dynamics of optically excited charge carriers in the antiferromagnetic semiconductor $\alpha $-MnTe at room temperature, by performing time-resolved 800 nm-pump terahertz-probe and 800 nm-pump 800 nm-probe spectroscopic measurements as a function of pump-probe delay for various pump fluences. Characteristic relaxation processes are identified not only in the low-fluence limit where a linear dependence occurs, but also tracked at high fluences beyond the linear regime. 
Our work motivates further investigation of the charge carrier dynamics deep into the antiferromagnetic phase.

\begin{acknowledgements}
We thank D. Kriegner and K. V\`{y}born\`{y} for helpful discussions. We acknowledge support by the European Research Council (ERC) under the Horizon 2020 research and innovation programme, Grant Agreement No. 950560 (DynaQuanta).
\end{acknowledgements}

\bibliographystyle{apsrev4-2}
\bibliography{MnTe_Paper_bib}

\begin{thebibliography}{27}%
\makeatletter
\providecommand \@ifxundefined [1]{%
 \@ifx{#1\undefined}
}%
\providecommand \@ifnum [1]{%
 \ifnum #1\expandafter \@firstoftwo
 \else \expandafter \@secondoftwo
 \fi
}%
\providecommand \@ifx [1]{%
 \ifx #1\expandafter \@firstoftwo
 \else \expandafter \@secondoftwo
 \fi
}%
\providecommand \natexlab [1]{#1}%
\providecommand \enquote  [1]{``#1''}%
\providecommand \bibnamefont  [1]{#1}%
\providecommand \bibfnamefont [1]{#1}%
\providecommand \citenamefont [1]{#1}%
\providecommand \href@noop [0]{\@secondoftwo}%
\providecommand \href [0]{\begingroup \@sanitize@url \@href}%
\providecommand \@href[1]{\@@startlink{#1}\@@href}%
\providecommand \@@href[1]{\endgroup#1\@@endlink}%
\providecommand \@sanitize@url [0]{\catcode `\\12\catcode `\$12\catcode
  `\&12\catcode `\#12\catcode `\^12\catcode `\_12\catcode `\%12\relax}%
\providecommand \@@startlink[1]{}%
\providecommand \@@endlink[0]{}%
\providecommand \url  [0]{\begingroup\@sanitize@url \@url }%
\providecommand \@url [1]{\endgroup\@href {#1}{\urlprefix }}%
\providecommand \urlprefix  [0]{URL }%
\providecommand \Eprint [0]{\href }%
\providecommand \doibase [0]{https://doi.org/}%
\providecommand \selectlanguage [0]{\@gobble}%
\providecommand \bibinfo  [0]{\@secondoftwo}%
\providecommand \bibfield  [0]{\@secondoftwo}%
\providecommand \translation [1]{[#1]}%
\providecommand \BibitemOpen [0]{}%
\providecommand \bibitemStop [0]{}%
\providecommand \bibitemNoStop [0]{.\EOS\space}%
\providecommand \EOS [0]{\spacefactor3000\relax}%
\providecommand \BibitemShut  [1]{\csname bibitem#1\endcsname}%
\let\auto@bib@innerbib\@empty
\bibitem [{\citenamefont {Jungwirth}\ \emph {et~al.}(2016)\citenamefont
  {Jungwirth}, \citenamefont {Marti}, \citenamefont {Wadley},\ and\
  \citenamefont {Wunderlich}}]{jungwirth2016antiferromagnetic}%
  \BibitemOpen
  \bibfield  {author} {\bibinfo {author} {\bibfnamefont {T.}~\bibnamefont
  {Jungwirth}}, \bibinfo {author} {\bibfnamefont {X.}~\bibnamefont {Marti}},
  \bibinfo {author} {\bibfnamefont {P.}~\bibnamefont {Wadley}},\ and\ \bibinfo
  {author} {\bibfnamefont {J.}~\bibnamefont {Wunderlich}},\ }\href@noop {}
  {\bibfield  {journal} {\bibinfo  {journal} {Nature nanotechnology}\ }\textbf
  {\bibinfo {volume} {11}},\ \bibinfo {pages} {231} (\bibinfo {year}
  {2016})}\BibitemShut {NoStop}%
\bibitem [{\citenamefont {Baltz}\ \emph {et~al.}(2018)\citenamefont {Baltz},
  \citenamefont {Manchon}, \citenamefont {Tsoi}, \citenamefont {Moriyama},
  \citenamefont {Ono},\ and\ \citenamefont
  {Tserkovnyak}}]{baltz2018antiferromagnetic}%
  \BibitemOpen
  \bibfield  {author} {\bibinfo {author} {\bibfnamefont {V.}~\bibnamefont
  {Baltz}}, \bibinfo {author} {\bibfnamefont {A.}~\bibnamefont {Manchon}},
  \bibinfo {author} {\bibfnamefont {M.}~\bibnamefont {Tsoi}}, \bibinfo {author}
  {\bibfnamefont {T.}~\bibnamefont {Moriyama}}, \bibinfo {author}
  {\bibfnamefont {T.}~\bibnamefont {Ono}},\ and\ \bibinfo {author}
  {\bibfnamefont {Y.}~\bibnamefont {Tserkovnyak}},\ }\href@noop {} {\bibfield
  {journal} {\bibinfo  {journal} {Reviews of Modern Physics}\ }\textbf
  {\bibinfo {volume} {90}},\ \bibinfo {pages} {015005} (\bibinfo {year}
  {2018})}\BibitemShut {NoStop}%
\bibitem [{\citenamefont {N{\v{e}}mec}\ \emph {et~al.}(2018)\citenamefont
  {N{\v{e}}mec}, \citenamefont {Fiebig}, \citenamefont {Kampfrath},\ and\
  \citenamefont {Kimel}}]{nvemec2018antiferromagnetic}%
  \BibitemOpen
  \bibfield  {author} {\bibinfo {author} {\bibfnamefont {P.}~\bibnamefont
  {N{\v{e}}mec}}, \bibinfo {author} {\bibfnamefont {M.}~\bibnamefont {Fiebig}},
  \bibinfo {author} {\bibfnamefont {T.}~\bibnamefont {Kampfrath}},\ and\
  \bibinfo {author} {\bibfnamefont {A.~V.}\ \bibnamefont {Kimel}},\ }\href@noop
  {} {\bibfield  {journal} {\bibinfo  {journal} {Nature Physics}\ }\textbf
  {\bibinfo {volume} {14}},\ \bibinfo {pages} {229} (\bibinfo {year}
  {2018})}\BibitemShut {NoStop}%
\bibitem [{\citenamefont {Kelley}(1939)}]{Kelley1939}%
  \BibitemOpen
  \bibfield  {author} {\bibinfo {author} {\bibfnamefont {K.~K.}\ \bibnamefont
  {Kelley}},\ }\href {https://doi.org/10.1021/ja01870a065} {\bibfield
  {journal} {\bibinfo  {journal} {Journal of the American Chemical Society}\
  }\textbf {\bibinfo {volume} {61}},\ \bibinfo {pages} {203} (\bibinfo {year}
  {1939})}\BibitemShut {NoStop}%
\bibitem [{\citenamefont {Kriegner}\ \emph {et~al.}(2016)\citenamefont
  {Kriegner}, \citenamefont {V{\`y}born{\`y}}, \citenamefont {Olejn{\'\i}k},
  \citenamefont {Reichlov{\'a}}, \citenamefont {Nov{\'a}k}, \citenamefont
  {Marti}, \citenamefont {Gazquez}, \citenamefont {Saidl}, \citenamefont
  {N{\v{e}}mec}, \citenamefont {Volobuev} \emph
  {et~al.}}]{kriegner2016multiple}%
  \BibitemOpen
  \bibfield  {author} {\bibinfo {author} {\bibfnamefont {D.}~\bibnamefont
  {Kriegner}}, \bibinfo {author} {\bibfnamefont {K.}~\bibnamefont
  {V{\`y}born{\`y}}}, \bibinfo {author} {\bibfnamefont {K.}~\bibnamefont
  {Olejn{\'\i}k}}, \bibinfo {author} {\bibfnamefont {H.}~\bibnamefont
  {Reichlov{\'a}}}, \bibinfo {author} {\bibfnamefont {V.}~\bibnamefont
  {Nov{\'a}k}}, \bibinfo {author} {\bibfnamefont {X.}~\bibnamefont {Marti}},
  \bibinfo {author} {\bibfnamefont {J.}~\bibnamefont {Gazquez}}, \bibinfo
  {author} {\bibfnamefont {V.}~\bibnamefont {Saidl}}, \bibinfo {author}
  {\bibfnamefont {P.}~\bibnamefont {N{\v{e}}mec}}, \bibinfo {author}
  {\bibfnamefont {V.}~\bibnamefont {Volobuev}}, \emph {et~al.},\ }\href@noop {}
  {\bibfield  {journal} {\bibinfo  {journal} {Nature communications}\ }\textbf
  {\bibinfo {volume} {7}},\ \bibinfo {pages} {11623} (\bibinfo {year}
  {2016})}\BibitemShut {NoStop}%
\bibitem [{\citenamefont {He}\ \emph {et~al.}(2018)\citenamefont {He},
  \citenamefont {Yin}, \citenamefont {Grutter}, \citenamefont {Pan},
  \citenamefont {Che}, \citenamefont {Yu}, \citenamefont {Gilbert},
  \citenamefont {Disseler}, \citenamefont {Liu}, \citenamefont {Shafer},
  \citenamefont {Zhang}, \citenamefont {Wu}, \citenamefont {Kirby},
  \citenamefont {Arenholz}, \citenamefont {Lake}, \citenamefont {Han},\ and\
  \citenamefont {Wang}}]{He2018}%
  \BibitemOpen
  \bibfield  {author} {\bibinfo {author} {\bibfnamefont {Q.~L.}\ \bibnamefont
  {He}}, \bibinfo {author} {\bibfnamefont {G.}~\bibnamefont {Yin}}, \bibinfo
  {author} {\bibfnamefont {A.~J.}\ \bibnamefont {Grutter}}, \bibinfo {author}
  {\bibfnamefont {L.}~\bibnamefont {Pan}}, \bibinfo {author} {\bibfnamefont
  {X.}~\bibnamefont {Che}}, \bibinfo {author} {\bibfnamefont {G.}~\bibnamefont
  {Yu}}, \bibinfo {author} {\bibfnamefont {D.~A.}\ \bibnamefont {Gilbert}},
  \bibinfo {author} {\bibfnamefont {S.~M.}\ \bibnamefont {Disseler}}, \bibinfo
  {author} {\bibfnamefont {Y.}~\bibnamefont {Liu}}, \bibinfo {author}
  {\bibfnamefont {P.}~\bibnamefont {Shafer}}, \bibinfo {author} {\bibfnamefont
  {B.}~\bibnamefont {Zhang}}, \bibinfo {author} {\bibfnamefont
  {Y.}~\bibnamefont {Wu}}, \bibinfo {author} {\bibfnamefont {B.~J.}\
  \bibnamefont {Kirby}}, \bibinfo {author} {\bibfnamefont {E.}~\bibnamefont
  {Arenholz}}, \bibinfo {author} {\bibfnamefont {R.~K.}\ \bibnamefont {Lake}},
  \bibinfo {author} {\bibfnamefont {X.}~\bibnamefont {Han}},\ and\ \bibinfo
  {author} {\bibfnamefont {K.~L.}\ \bibnamefont {Wang}},\ }\href
  {https://doi.org/10.1038/s41467-018-05166-9} {\bibfield  {journal} {\bibinfo
  {journal} {Nature Communications}\ }\textbf {\bibinfo {volume} {9}},\
  \bibinfo {pages} {2767} (\bibinfo {year} {2018})}\BibitemShut {NoStop}%
\bibitem [{\citenamefont {Zheng}\ \emph {et~al.}(2019)\citenamefont {Zheng},
  \citenamefont {Lu}, \citenamefont {Polash}, \citenamefont
  {Rasoulianboroujeni}, \citenamefont {Liu}, \citenamefont {Manley},
  \citenamefont {Deng}, \citenamefont {Sun}, \citenamefont {Chen},
  \citenamefont {Hermann}, \citenamefont {Vashaee}, \citenamefont {Heremans},\
  and\ \citenamefont {Zhao}}]{Zheng19}%
  \BibitemOpen
  \bibfield  {author} {\bibinfo {author} {\bibfnamefont {Y.}~\bibnamefont
  {Zheng}}, \bibinfo {author} {\bibfnamefont {T.}~\bibnamefont {Lu}}, \bibinfo
  {author} {\bibfnamefont {M.~M.~H.}\ \bibnamefont {Polash}}, \bibinfo {author}
  {\bibfnamefont {M.}~\bibnamefont {Rasoulianboroujeni}}, \bibinfo {author}
  {\bibfnamefont {N.}~\bibnamefont {Liu}}, \bibinfo {author} {\bibfnamefont
  {M.~E.}\ \bibnamefont {Manley}}, \bibinfo {author} {\bibfnamefont
  {Y.}~\bibnamefont {Deng}}, \bibinfo {author} {\bibfnamefont {P.~J.}\
  \bibnamefont {Sun}}, \bibinfo {author} {\bibfnamefont {X.~L.}\ \bibnamefont
  {Chen}}, \bibinfo {author} {\bibfnamefont {R.~P.}\ \bibnamefont {Hermann}},
  \bibinfo {author} {\bibfnamefont {D.}~\bibnamefont {Vashaee}}, \bibinfo
  {author} {\bibfnamefont {J.~P.}\ \bibnamefont {Heremans}},\ and\ \bibinfo
  {author} {\bibfnamefont {H.}~\bibnamefont {Zhao}},\ }\href
  {https://doi.org/10.1126/sciadv.aat9461} {\bibfield  {journal} {\bibinfo
  {journal} {Science Advances}\ }\textbf {\bibinfo {volume} {5}},\ \bibinfo
  {pages} {eaat9461} (\bibinfo {year} {2019})}\BibitemShut {NoStop}%
\bibitem [{\citenamefont {Bossini}\ \emph {et~al.}(2020)\citenamefont
  {Bossini}, \citenamefont {Terschanski}, \citenamefont {Mertens},
  \citenamefont {Springholz}, \citenamefont {Bonanni}, \citenamefont {Uhrig},\
  and\ \citenamefont {Cinchetti}}]{bossini2020exchange}%
  \BibitemOpen
  \bibfield  {author} {\bibinfo {author} {\bibfnamefont {D.}~\bibnamefont
  {Bossini}}, \bibinfo {author} {\bibfnamefont {M.}~\bibnamefont
  {Terschanski}}, \bibinfo {author} {\bibfnamefont {F.}~\bibnamefont
  {Mertens}}, \bibinfo {author} {\bibfnamefont {G.}~\bibnamefont {Springholz}},
  \bibinfo {author} {\bibfnamefont {A.}~\bibnamefont {Bonanni}}, \bibinfo
  {author} {\bibfnamefont {G.~S.}\ \bibnamefont {Uhrig}},\ and\ \bibinfo
  {author} {\bibfnamefont {M.}~\bibnamefont {Cinchetti}},\ }\href@noop {}
  {\bibfield  {journal} {\bibinfo  {journal} {New Journal of Physics}\ }\textbf
  {\bibinfo {volume} {22}},\ \bibinfo {pages} {083029} (\bibinfo {year}
  {2020})}\BibitemShut {NoStop}%
\bibitem [{\citenamefont {Bossini}\ \emph {et~al.}(2021)\citenamefont
  {Bossini}, \citenamefont {Dal~Conte}, \citenamefont {Terschanski},
  \citenamefont {Springholz}, \citenamefont {Bonanni}, \citenamefont
  {Deltenre}, \citenamefont {Anders}, \citenamefont {Uhrig}, \citenamefont
  {Cerullo},\ and\ \citenamefont {Cinchetti}}]{bossini2021femtosecond}%
  \BibitemOpen
  \bibfield  {author} {\bibinfo {author} {\bibfnamefont {D.}~\bibnamefont
  {Bossini}}, \bibinfo {author} {\bibfnamefont {S.}~\bibnamefont {Dal~Conte}},
  \bibinfo {author} {\bibfnamefont {M.}~\bibnamefont {Terschanski}}, \bibinfo
  {author} {\bibfnamefont {G.}~\bibnamefont {Springholz}}, \bibinfo {author}
  {\bibfnamefont {A.}~\bibnamefont {Bonanni}}, \bibinfo {author} {\bibfnamefont
  {K.}~\bibnamefont {Deltenre}}, \bibinfo {author} {\bibfnamefont
  {F.}~\bibnamefont {Anders}}, \bibinfo {author} {\bibfnamefont {G.~S.}\
  \bibnamefont {Uhrig}}, \bibinfo {author} {\bibfnamefont {G.}~\bibnamefont
  {Cerullo}},\ and\ \bibinfo {author} {\bibfnamefont {M.}~\bibnamefont
  {Cinchetti}},\ }\href@noop {} {\bibfield  {journal} {\bibinfo  {journal}
  {Physical Review B}\ }\textbf {\bibinfo {volume} {104}},\ \bibinfo {pages}
  {224424} (\bibinfo {year} {2021})}\BibitemShut {NoStop}%
\bibitem [{\citenamefont {Wang}\ \emph {et~al.}(2021)\citenamefont {Wang},
  \citenamefont {Yang}, \citenamefont {Zhang}, \citenamefont {Zhou},
  \citenamefont {Wang}, \citenamefont {Hu}, \citenamefont {Xue},\ and\
  \citenamefont {Xu}}]{Wang2021}%
  \BibitemOpen
  \bibfield  {author} {\bibinfo {author} {\bibfnamefont {F.}~\bibnamefont
  {Wang}}, \bibinfo {author} {\bibfnamefont {H.}~\bibnamefont {Yang}}, \bibinfo
  {author} {\bibfnamefont {H.}~\bibnamefont {Zhang}}, \bibinfo {author}
  {\bibfnamefont {J.}~\bibnamefont {Zhou}}, \bibinfo {author} {\bibfnamefont
  {J.}~\bibnamefont {Wang}}, \bibinfo {author} {\bibfnamefont {L.}~\bibnamefont
  {Hu}}, \bibinfo {author} {\bibfnamefont {D.-J.}\ \bibnamefont {Xue}},\ and\
  \bibinfo {author} {\bibfnamefont {X.}~\bibnamefont {Xu}},\ }\href
  {https://doi.org/10.1021/acs.nanolett.1c02481} {\bibfield  {journal}
  {\bibinfo  {journal} {Nano Letters}\ }\textbf {\bibinfo {volume} {21}},\
  \bibinfo {pages} {7684} (\bibinfo {year} {2021})}\BibitemShut {NoStop}%
\bibitem [{\citenamefont {Gonzalez~Betancourt}\ \emph
  {et~al.}(2023)\citenamefont {Gonzalez~Betancourt}, \citenamefont
  {Zub\'a\ifmmode~\check{c}\else \v{c}\fi{}}, \citenamefont
  {Gonzalez-Hernandez}, \citenamefont {Geishendorf}, \citenamefont {\ifmmode
  \check{S}\else \v{S}\fi{}ob\'a\ifmmode~\check{n}\else \v{n}\fi{}},
  \citenamefont {Springholz}, \citenamefont {Olejn\'{\i}k}, \citenamefont
  {\ifmmode~\check{S}\else \v{S}\fi{}mejkal}, \citenamefont {Sinova},
  \citenamefont {Jungwirth}, \citenamefont {Goennenwein}, \citenamefont
  {Thomas}, \citenamefont {Reichlov\'a}, \citenamefont {\ifmmode~\check{Z}\else
  \v{Z}\fi{}elezn\'y},\ and\ \citenamefont
  {Kriegner}}]{PhysRevLett.130.036702}%
  \BibitemOpen
  \bibfield  {author} {\bibinfo {author} {\bibfnamefont {R.~D.}\ \bibnamefont
  {Gonzalez~Betancourt}}, \bibinfo {author} {\bibfnamefont {J.}~\bibnamefont
  {Zub\'a\ifmmode~\check{c}\else \v{c}\fi{}}}, \bibinfo {author} {\bibfnamefont
  {R.}~\bibnamefont {Gonzalez-Hernandez}}, \bibinfo {author} {\bibfnamefont
  {K.}~\bibnamefont {Geishendorf}}, \bibinfo {author} {\bibfnamefont
  {Z.}~\bibnamefont {\ifmmode \check{S}\else
  \v{S}\fi{}ob\'a\ifmmode~\check{n}\else \v{n}\fi{}}}, \bibinfo {author}
  {\bibfnamefont {G.}~\bibnamefont {Springholz}}, \bibinfo {author}
  {\bibfnamefont {K.}~\bibnamefont {Olejn\'{\i}k}}, \bibinfo {author}
  {\bibfnamefont {L.}~\bibnamefont {\ifmmode~\check{S}\else \v{S}\fi{}mejkal}},
  \bibinfo {author} {\bibfnamefont {J.}~\bibnamefont {Sinova}}, \bibinfo
  {author} {\bibfnamefont {T.}~\bibnamefont {Jungwirth}}, \bibinfo {author}
  {\bibfnamefont {S.~T.~B.}\ \bibnamefont {Goennenwein}}, \bibinfo {author}
  {\bibfnamefont {A.}~\bibnamefont {Thomas}}, \bibinfo {author} {\bibfnamefont
  {H.}~\bibnamefont {Reichlov\'a}}, \bibinfo {author} {\bibfnamefont
  {J.}~\bibnamefont {\ifmmode~\check{Z}\else \v{Z}\fi{}elezn\'y}},\ and\
  \bibinfo {author} {\bibfnamefont {D.}~\bibnamefont {Kriegner}},\ }\href
  {https://doi.org/10.1103/PhysRevLett.130.036702} {\bibfield  {journal}
  {\bibinfo  {journal} {Phys. Rev. Lett.}\ }\textbf {\bibinfo {volume} {130}},\
  \bibinfo {pages} {036702} (\bibinfo {year} {2023})}\BibitemShut {NoStop}%
\bibitem [{\citenamefont {Moseley}\ \emph {et~al.}(2022)\citenamefont
  {Moseley}, \citenamefont {Taddei}, \citenamefont {Yan}, \citenamefont
  {McGuire}, \citenamefont {Calder}, \citenamefont {Polash}, \citenamefont
  {Vashaee}, \citenamefont {Zhang}, \citenamefont {Zhao}, \citenamefont
  {Parker} \emph {et~al.}}]{moseley2022giant}%
  \BibitemOpen
  \bibfield  {author} {\bibinfo {author} {\bibfnamefont {D.~H.}\ \bibnamefont
  {Moseley}}, \bibinfo {author} {\bibfnamefont {K.~M.}\ \bibnamefont {Taddei}},
  \bibinfo {author} {\bibfnamefont {J.}~\bibnamefont {Yan}}, \bibinfo {author}
  {\bibfnamefont {M.~A.}\ \bibnamefont {McGuire}}, \bibinfo {author}
  {\bibfnamefont {S.}~\bibnamefont {Calder}}, \bibinfo {author} {\bibfnamefont
  {M.~M.~H.}\ \bibnamefont {Polash}}, \bibinfo {author} {\bibfnamefont
  {D.}~\bibnamefont {Vashaee}}, \bibinfo {author} {\bibfnamefont
  {X.}~\bibnamefont {Zhang}}, \bibinfo {author} {\bibfnamefont
  {H.}~\bibnamefont {Zhao}}, \bibinfo {author} {\bibfnamefont {D.~S.}\
  \bibnamefont {Parker}}, \emph {et~al.},\ }\href@noop {} {\bibfield  {journal}
  {\bibinfo  {journal} {Physical Review Materials}\ }\textbf {\bibinfo {volume}
  {6}},\ \bibinfo {pages} {014404} (\bibinfo {year} {2022})}\BibitemShut
  {NoStop}%
\bibitem [{\citenamefont {Yao}\ \emph {et~al.}(2022)\citenamefont {Yao},
  \citenamefont {Mazza}, \citenamefont {Han}, \citenamefont {Yi}, \citenamefont
  {Jain}, \citenamefont {Brahlek},\ and\ \citenamefont {Oh}}]{Yao2022}%
  \BibitemOpen
  \bibfield  {author} {\bibinfo {author} {\bibfnamefont {X.}~\bibnamefont
  {Yao}}, \bibinfo {author} {\bibfnamefont {A.~R.}\ \bibnamefont {Mazza}},
  \bibinfo {author} {\bibfnamefont {M.-G.}\ \bibnamefont {Han}}, \bibinfo
  {author} {\bibfnamefont {H.~T.}\ \bibnamefont {Yi}}, \bibinfo {author}
  {\bibfnamefont {D.}~\bibnamefont {Jain}}, \bibinfo {author} {\bibfnamefont
  {M.}~\bibnamefont {Brahlek}},\ and\ \bibinfo {author} {\bibfnamefont
  {S.}~\bibnamefont {Oh}},\ }\href
  {https://doi.org/10.1021/acs.nanolett.2c02510} {\bibfield  {journal}
  {\bibinfo  {journal} {Nano Letters}\ }\textbf {\bibinfo {volume} {22}},\
  \bibinfo {pages} {7522} (\bibinfo {year} {2022})}\BibitemShut {NoStop}%
\bibitem [{\citenamefont {{Efrem D'Sa}}\ \emph {et~al.}(2005)\citenamefont
  {{Efrem D'Sa}}, \citenamefont {Bhobe}, \citenamefont {Priolkar},
  \citenamefont {Das}, \citenamefont {Paranjpe}, \citenamefont {Prabhu},\ and\
  \citenamefont {Sarode}}]{EFREMDSA2005267}%
  \BibitemOpen
  \bibfield  {author} {\bibinfo {author} {\bibfnamefont {J.}~\bibnamefont
  {{Efrem D'Sa}}}, \bibinfo {author} {\bibfnamefont {P.}~\bibnamefont {Bhobe}},
  \bibinfo {author} {\bibfnamefont {K.}~\bibnamefont {Priolkar}}, \bibinfo
  {author} {\bibfnamefont {A.}~\bibnamefont {Das}}, \bibinfo {author}
  {\bibfnamefont {S.}~\bibnamefont {Paranjpe}}, \bibinfo {author}
  {\bibfnamefont {R.}~\bibnamefont {Prabhu}},\ and\ \bibinfo {author}
  {\bibfnamefont {P.}~\bibnamefont {Sarode}},\ }\href
  {https://doi.org/https://doi.org/10.1016/j.jmmm.2004.08.001} {\bibfield
  {journal} {\bibinfo  {journal} {Journal of Magnetism and Magnetic Materials}\
  }\textbf {\bibinfo {volume} {285}},\ \bibinfo {pages} {267} (\bibinfo {year}
  {2005})}\BibitemShut {NoStop}%
\bibitem [{\citenamefont {{Kunitomi, Nobuhiko}}\ \emph
  {et~al.}(1964)\citenamefont {{Kunitomi, Nobuhiko}}, \citenamefont
  {{Hamaguchi, Yoshikazu}},\ and\ \citenamefont {{Anzai,
  Shuichiro}}}]{Kunitomi64}%
  \BibitemOpen
  \bibfield  {author} {\bibinfo {author} {\bibnamefont {{Kunitomi, Nobuhiko}}},
  \bibinfo {author} {\bibnamefont {{Hamaguchi, Yoshikazu}}},\ and\ \bibinfo
  {author} {\bibnamefont {{Anzai, Shuichiro}}},\ }\href
  {https://doi.org/10.1051/jphys:01964002505056800} {\bibfield  {journal}
  {\bibinfo  {journal} {J. Phys. France}\ }\textbf {\bibinfo {volume} {25}},\
  \bibinfo {pages} {568} (\bibinfo {year} {1964})}\BibitemShut {NoStop}%
\bibitem [{\citenamefont {Onari}\ \emph {et~al.}(1974)\citenamefont {Onari},
  \citenamefont {Arai},\ and\ \citenamefont {Kudo}}]{onari1974temperature}%
  \BibitemOpen
  \bibfield  {author} {\bibinfo {author} {\bibfnamefont {S.}~\bibnamefont
  {Onari}}, \bibinfo {author} {\bibfnamefont {T.}~\bibnamefont {Arai}},\ and\
  \bibinfo {author} {\bibfnamefont {K.}~\bibnamefont {Kudo}},\ }\href@noop {}
  {\bibfield  {journal} {\bibinfo  {journal} {Solid State Communications}\
  }\textbf {\bibinfo {volume} {14}},\ \bibinfo {pages} {507} (\bibinfo {year}
  {1974})}\BibitemShut {NoStop}%
\bibitem [{\citenamefont {Ferrer-Roca}\ \emph {et~al.}(2000)\citenamefont
  {Ferrer-Roca}, \citenamefont {Segura}, \citenamefont {Reig},\ and\
  \citenamefont {Mu\~noz}}]{Ferrer00}%
  \BibitemOpen
  \bibfield  {author} {\bibinfo {author} {\bibfnamefont {C.}~\bibnamefont
  {Ferrer-Roca}}, \bibinfo {author} {\bibfnamefont {A.}~\bibnamefont {Segura}},
  \bibinfo {author} {\bibfnamefont {C.}~\bibnamefont {Reig}},\ and\ \bibinfo
  {author} {\bibfnamefont {V.}~\bibnamefont {Mu\~noz}},\ }\href
  {https://doi.org/10.1103/PhysRevB.61.13679} {\bibfield  {journal} {\bibinfo
  {journal} {Phys. Rev. B}\ }\textbf {\bibinfo {volume} {61}},\ \bibinfo
  {pages} {13679} (\bibinfo {year} {2000})}\BibitemShut {NoStop}%
\bibitem [{\citenamefont {Wei}\ and\ \citenamefont {Zunger}(1987)}]{Wei87}%
  \BibitemOpen
  \bibfield  {author} {\bibinfo {author} {\bibfnamefont {S.-H.}\ \bibnamefont
  {Wei}}\ and\ \bibinfo {author} {\bibfnamefont {A.}~\bibnamefont {Zunger}},\
  }\href {https://doi.org/10.1103/PhysRevB.35.2340} {\bibfield  {journal}
  {\bibinfo  {journal} {Phys. Rev. B}\ }\textbf {\bibinfo {volume} {35}},\
  \bibinfo {pages} {2340} (\bibinfo {year} {1987})}\BibitemShut {NoStop}%
\bibitem [{\citenamefont {Przezdziecka}\ \emph {et~al.}(2005)\citenamefont
  {Przezdziecka}, \citenamefont {Kaminska}, \citenamefont {Dynowska},
  \citenamefont {Butkute}, \citenamefont {Dobrowolski}, \citenamefont {Kepa},
  \citenamefont {Jakiela}, \citenamefont {Aleszkiewicz}, \citenamefont
  {Lusakowska}, \citenamefont {Janik},\ and\ \citenamefont
  {Kossut}}]{Przezdziecka05}%
  \BibitemOpen
  \bibfield  {author} {\bibinfo {author} {\bibfnamefont {E.}~\bibnamefont
  {Przezdziecka}}, \bibinfo {author} {\bibfnamefont {E.}~\bibnamefont
  {Kaminska}}, \bibinfo {author} {\bibfnamefont {E.}~\bibnamefont {Dynowska}},
  \bibinfo {author} {\bibfnamefont {R.}~\bibnamefont {Butkute}}, \bibinfo
  {author} {\bibfnamefont {W.}~\bibnamefont {Dobrowolski}}, \bibinfo {author}
  {\bibfnamefont {H.}~\bibnamefont {Kepa}}, \bibinfo {author} {\bibfnamefont
  {R.}~\bibnamefont {Jakiela}}, \bibinfo {author} {\bibfnamefont
  {M.}~\bibnamefont {Aleszkiewicz}}, \bibinfo {author} {\bibfnamefont
  {E.}~\bibnamefont {Lusakowska}}, \bibinfo {author} {\bibfnamefont
  {E.}~\bibnamefont {Janik}},\ and\ \bibinfo {author} {\bibfnamefont
  {J.}~\bibnamefont {Kossut}},\ }\href
  {https://doi.org/https://doi.org/10.1002/pssc.200460667} {\bibfield
  {journal} {\bibinfo  {journal} {physica status solidi (c)}\ }\textbf
  {\bibinfo {volume} {2}},\ \bibinfo {pages} {1218} (\bibinfo {year}
  {2005})}\BibitemShut {NoStop}%
\bibitem [{\citenamefont {Hebling}\ \emph {et~al.}(2002)\citenamefont
  {Hebling}, \citenamefont {Almasi}, \citenamefont {Kozma},\ and\ \citenamefont
  {Kuhl}}]{Hebling02}%
  \BibitemOpen
  \bibfield  {author} {\bibinfo {author} {\bibfnamefont {J.}~\bibnamefont
  {Hebling}}, \bibinfo {author} {\bibfnamefont {G.}~\bibnamefont {Almasi}},
  \bibinfo {author} {\bibfnamefont {I.~Z.}\ \bibnamefont {Kozma}},\ and\
  \bibinfo {author} {\bibfnamefont {J.}~\bibnamefont {Kuhl}},\ }\href@noop {}
  {\bibfield  {journal} {\bibinfo  {journal} {Optics Express}\ }\textbf
  {\bibinfo {volume} {10}},\ \bibinfo {pages} {1161} (\bibinfo {year}
  {2002})}\BibitemShut {NoStop}%
\bibitem [{\citenamefont {Planken}\ \emph {et~al.}(2001)\citenamefont
  {Planken}, \citenamefont {Nienhuys}, \citenamefont {Bakker},\ and\
  \citenamefont {Wenckebach}}]{Planken01}%
  \BibitemOpen
  \bibfield  {author} {\bibinfo {author} {\bibfnamefont {P.~C.~M.}\
  \bibnamefont {Planken}}, \bibinfo {author} {\bibfnamefont {H.-K.}\
  \bibnamefont {Nienhuys}}, \bibinfo {author} {\bibfnamefont {H.~J.}\
  \bibnamefont {Bakker}},\ and\ \bibinfo {author} {\bibfnamefont
  {T.}~\bibnamefont {Wenckebach}},\ }\href
  {https://doi.org/10.1364/JOSAB.18.000313} {\bibfield  {journal} {\bibinfo
  {journal} {J. Opt. Soc. Am. B}\ }\textbf {\bibinfo {volume} {18}},\ \bibinfo
  {pages} {313} (\bibinfo {year} {2001})}\BibitemShut {NoStop}%
\bibitem [{\citenamefont {Reinhoffer}\ \emph {et~al.}(2020)\citenamefont
  {Reinhoffer}, \citenamefont {Mukai}, \citenamefont {Germanskiy},
  \citenamefont {Bliesener}, \citenamefont {Lippertz}, \citenamefont {Uday},
  \citenamefont {Taskin}, \citenamefont {Ando}, \citenamefont {Wang},\ and\
  \citenamefont {van Loosdrecht}}]{Reinhoffer20}%
  \BibitemOpen
  \bibfield  {author} {\bibinfo {author} {\bibfnamefont {C.}~\bibnamefont
  {Reinhoffer}}, \bibinfo {author} {\bibfnamefont {Y.}~\bibnamefont {Mukai}},
  \bibinfo {author} {\bibfnamefont {S.}~\bibnamefont {Germanskiy}}, \bibinfo
  {author} {\bibfnamefont {A.}~\bibnamefont {Bliesener}}, \bibinfo {author}
  {\bibfnamefont {G.}~\bibnamefont {Lippertz}}, \bibinfo {author}
  {\bibfnamefont {A.}~\bibnamefont {Uday}}, \bibinfo {author} {\bibfnamefont
  {A.~A.}\ \bibnamefont {Taskin}}, \bibinfo {author} {\bibfnamefont
  {Y.}~\bibnamefont {Ando}}, \bibinfo {author} {\bibfnamefont {Z.}~\bibnamefont
  {Wang}},\ and\ \bibinfo {author} {\bibfnamefont {P.~H.~M.}\ \bibnamefont {van
  Loosdrecht}},\ }\href {https://doi.org/10.1103/PhysRevMaterials.4.124201}
  {\bibfield  {journal} {\bibinfo  {journal} {Phys. Rev. Mater.}\ }\textbf
  {\bibinfo {volume} {4}},\ \bibinfo {pages} {124201} (\bibinfo {year}
  {2020})}\BibitemShut {NoStop}%
\bibitem [{\citenamefont {Kovalev}\ \emph {et~al.}(2020)\citenamefont
  {Kovalev}, \citenamefont {Dantas}, \citenamefont {Germanskiy}, \citenamefont
  {Deinert}, \citenamefont {Green}, \citenamefont {Ilyakov}, \citenamefont
  {Awari}, \citenamefont {Chen}, \citenamefont {Bawatna}, \citenamefont {Ling},
  \citenamefont {Xiu}, \citenamefont {van Loosdrecht}, \citenamefont
  {Surówka}, \citenamefont {Oka},\ and\ \citenamefont {Wang}}]{Kovalev20}%
  \BibitemOpen
  \bibfield  {author} {\bibinfo {author} {\bibfnamefont {S.}~\bibnamefont
  {Kovalev}}, \bibinfo {author} {\bibfnamefont {R.~M.~A.}\ \bibnamefont
  {Dantas}}, \bibinfo {author} {\bibfnamefont {S.}~\bibnamefont {Germanskiy}},
  \bibinfo {author} {\bibfnamefont {J.-C.}\ \bibnamefont {Deinert}}, \bibinfo
  {author} {\bibfnamefont {B.}~\bibnamefont {Green}}, \bibinfo {author}
  {\bibfnamefont {I.}~\bibnamefont {Ilyakov}}, \bibinfo {author} {\bibfnamefont
  {N.}~\bibnamefont {Awari}}, \bibinfo {author} {\bibfnamefont
  {M.}~\bibnamefont {Chen}}, \bibinfo {author} {\bibfnamefont {M.}~\bibnamefont
  {Bawatna}}, \bibinfo {author} {\bibfnamefont {J.}~\bibnamefont {Ling}},
  \bibinfo {author} {\bibfnamefont {F.}~\bibnamefont {Xiu}}, \bibinfo {author}
  {\bibfnamefont {P.~H.~M.}\ \bibnamefont {van Loosdrecht}}, \bibinfo {author}
  {\bibfnamefont {P.}~\bibnamefont {Surówka}}, \bibinfo {author}
  {\bibfnamefont {T.}~\bibnamefont {Oka}},\ and\ \bibinfo {author}
  {\bibfnamefont {Z.}~\bibnamefont {Wang}},\ }\href
  {https://doi.org/10.1038/s41467-020-16133-8} {\bibfield  {journal} {\bibinfo
  {journal} {Nature Commun.}\ }\textbf {\bibinfo {volume} {11}},\ \bibinfo
  {pages} {2451} (\bibinfo {year} {2020})}\BibitemShut {NoStop}%
\bibitem [{\citenamefont {Kovalev}\ \emph {et~al.}(2021)\citenamefont
  {Kovalev}, \citenamefont {Dong}, \citenamefont {Shi}, \citenamefont
  {Reinhoffer}, \citenamefont {Xu}, \citenamefont {Wang}, \citenamefont {Wang},
  \citenamefont {Gan}, \citenamefont {Germanskiy}, \citenamefont {Deinert},
  \citenamefont {Ilyakov}, \citenamefont {van Loosdrecht}, \citenamefont {Wu},
  \citenamefont {Wang}, \citenamefont {Demsar},\ and\ \citenamefont
  {Wang}}]{Kovalev21}%
  \BibitemOpen
  \bibfield  {author} {\bibinfo {author} {\bibfnamefont {S.}~\bibnamefont
  {Kovalev}}, \bibinfo {author} {\bibfnamefont {T.}~\bibnamefont {Dong}},
  \bibinfo {author} {\bibfnamefont {L.-Y.}\ \bibnamefont {Shi}}, \bibinfo
  {author} {\bibfnamefont {C.}~\bibnamefont {Reinhoffer}}, \bibinfo {author}
  {\bibfnamefont {T.-Q.}\ \bibnamefont {Xu}}, \bibinfo {author} {\bibfnamefont
  {H.-Z.}\ \bibnamefont {Wang}}, \bibinfo {author} {\bibfnamefont
  {Y.}~\bibnamefont {Wang}}, \bibinfo {author} {\bibfnamefont {Z.-Z.}\
  \bibnamefont {Gan}}, \bibinfo {author} {\bibfnamefont {S.}~\bibnamefont
  {Germanskiy}}, \bibinfo {author} {\bibfnamefont {J.-C.}\ \bibnamefont
  {Deinert}}, \bibinfo {author} {\bibfnamefont {I.}~\bibnamefont {Ilyakov}},
  \bibinfo {author} {\bibfnamefont {P.~H.~M.}\ \bibnamefont {van Loosdrecht}},
  \bibinfo {author} {\bibfnamefont {D.}~\bibnamefont {Wu}}, \bibinfo {author}
  {\bibfnamefont {N.-L.}\ \bibnamefont {Wang}}, \bibinfo {author}
  {\bibfnamefont {J.}~\bibnamefont {Demsar}},\ and\ \bibinfo {author}
  {\bibfnamefont {Z.}~\bibnamefont {Wang}},\ }\href
  {https://doi.org/10.1103/PhysRevB.104.L140505} {\bibfield  {journal}
  {\bibinfo  {journal} {Phys. Rev. B}\ }\textbf {\bibinfo {volume} {104}},\
  \bibinfo {pages} {L140505} (\bibinfo {year} {2021})}\BibitemShut {NoStop}%
\bibitem [{\citenamefont {Germanskiy}\ \emph {et~al.}(2022)\citenamefont
  {Germanskiy}, \citenamefont {Dantas}, \citenamefont {Kovalev}, \citenamefont
  {Reinhoffer}, \citenamefont {Mashkovich}, \citenamefont {van Loosdrecht},
  \citenamefont {Yang}, \citenamefont {Xiu}, \citenamefont {Sur\'owka},
  \citenamefont {Moessner}, \citenamefont {Oka},\ and\ \citenamefont
  {Wang}}]{Germanskiy22}%
  \BibitemOpen
  \bibfield  {author} {\bibinfo {author} {\bibfnamefont {S.}~\bibnamefont
  {Germanskiy}}, \bibinfo {author} {\bibfnamefont {R.~M.~A.}\ \bibnamefont
  {Dantas}}, \bibinfo {author} {\bibfnamefont {S.}~\bibnamefont {Kovalev}},
  \bibinfo {author} {\bibfnamefont {C.}~\bibnamefont {Reinhoffer}}, \bibinfo
  {author} {\bibfnamefont {E.~A.}\ \bibnamefont {Mashkovich}}, \bibinfo
  {author} {\bibfnamefont {P.~H.~M.}\ \bibnamefont {van Loosdrecht}}, \bibinfo
  {author} {\bibfnamefont {Y.}~\bibnamefont {Yang}}, \bibinfo {author}
  {\bibfnamefont {F.}~\bibnamefont {Xiu}}, \bibinfo {author} {\bibfnamefont
  {P.}~\bibnamefont {Sur\'owka}}, \bibinfo {author} {\bibfnamefont
  {R.}~\bibnamefont {Moessner}}, \bibinfo {author} {\bibfnamefont
  {T.}~\bibnamefont {Oka}},\ and\ \bibinfo {author} {\bibfnamefont
  {Z.}~\bibnamefont {Wang}},\ }\href
  {https://doi.org/10.1103/PhysRevB.106.L081127} {\bibfield  {journal}
  {\bibinfo  {journal} {Phys. Rev. B}\ }\textbf {\bibinfo {volume} {106}},\
  \bibinfo {pages} {L081127} (\bibinfo {year} {2022})}\BibitemShut {NoStop}%
\bibitem [{\citenamefont {Reinhoffer}\ \emph {et~al.}(2022)\citenamefont
  {Reinhoffer}, \citenamefont {Pilch}, \citenamefont {Reinold}, \citenamefont
  {Derendorf}, \citenamefont {Kovalev}, \citenamefont {Deinert}, \citenamefont
  {Ilyakov}, \citenamefont {Ponomaryov}, \citenamefont {Chen}, \citenamefont
  {Xu}, \citenamefont {Wang}, \citenamefont {Gan}, \citenamefont {Wu},
  \citenamefont {Luo}, \citenamefont {Germanskiy}, \citenamefont {Mashkovich},
  \citenamefont {van Loosdrecht}, \citenamefont {Eremin},\ and\ \citenamefont
  {Wang}}]{Reinhoffer22}%
  \BibitemOpen
  \bibfield  {author} {\bibinfo {author} {\bibfnamefont {C.}~\bibnamefont
  {Reinhoffer}}, \bibinfo {author} {\bibfnamefont {P.}~\bibnamefont {Pilch}},
  \bibinfo {author} {\bibfnamefont {A.}~\bibnamefont {Reinold}}, \bibinfo
  {author} {\bibfnamefont {P.}~\bibnamefont {Derendorf}}, \bibinfo {author}
  {\bibfnamefont {S.}~\bibnamefont {Kovalev}}, \bibinfo {author} {\bibfnamefont
  {J.-C.}\ \bibnamefont {Deinert}}, \bibinfo {author} {\bibfnamefont
  {I.}~\bibnamefont {Ilyakov}}, \bibinfo {author} {\bibfnamefont
  {A.}~\bibnamefont {Ponomaryov}}, \bibinfo {author} {\bibfnamefont
  {M.}~\bibnamefont {Chen}}, \bibinfo {author} {\bibfnamefont {T.-Q.}\
  \bibnamefont {Xu}}, \bibinfo {author} {\bibfnamefont {Y.}~\bibnamefont
  {Wang}}, \bibinfo {author} {\bibfnamefont {Z.-Z.}\ \bibnamefont {Gan}},
  \bibinfo {author} {\bibfnamefont {D.-S.}\ \bibnamefont {Wu}}, \bibinfo
  {author} {\bibfnamefont {J.-L.}\ \bibnamefont {Luo}}, \bibinfo {author}
  {\bibfnamefont {S.}~\bibnamefont {Germanskiy}}, \bibinfo {author}
  {\bibfnamefont {E.~A.}\ \bibnamefont {Mashkovich}}, \bibinfo {author}
  {\bibfnamefont {P.~H.~M.}\ \bibnamefont {van Loosdrecht}}, \bibinfo {author}
  {\bibfnamefont {I.~M.}\ \bibnamefont {Eremin}},\ and\ \bibinfo {author}
  {\bibfnamefont {Z.}~\bibnamefont {Wang}},\ }\href
  {https://doi.org/10.1103/PhysRevB.106.214514} {\bibfield  {journal} {\bibinfo
   {journal} {Phys. Rev. B}\ }\textbf {\bibinfo {volume} {106}},\ \bibinfo
  {pages} {214514} (\bibinfo {year} {2022})}\BibitemShut {NoStop}%
\bibitem [{\citenamefont {Myasnikova}\ \emph {et~al.}(2020)\citenamefont
  {Myasnikova}, \citenamefont {Shendrik},\ and\ \citenamefont
  {Bogdanov}}]{Myasnikova20}%
  \BibitemOpen
  \bibfield  {author} {\bibinfo {author} {\bibfnamefont {A.}~\bibnamefont
  {Myasnikova}}, \bibinfo {author} {\bibfnamefont {R.}~\bibnamefont
  {Shendrik}},\ and\ \bibinfo {author} {\bibfnamefont {A.}~\bibnamefont
  {Bogdanov}},\ }\href {https://doi.org/10.1039/D0RA00865F} {\bibfield
  {journal} {\bibinfo  {journal} {RSC Adv.}\ }\textbf {\bibinfo {volume}
  {10}},\ \bibinfo {pages} {13992} (\bibinfo {year} {2020})}\BibitemShut
  {NoStop}%
\end{thebibliography}%

\end{document}